\documentclass[aps,prb,reprint,superscriptaddress]{revtex4}

\usepackage{graphicx}
\usepackage{dcolumn}
\usepackage{bm}
\usepackage{amsmath}
\usepackage{amssymb}
\usepackage{latexsym}
\usepackage{epsfig}
\usepackage{amsbsy}
\usepackage{array}
\usepackage{amssymb}
\usepackage{setspace}
\usepackage{bm}
\usepackage[colorlinks=true, linkcolor=blue, citecolor=blue, urlcolor=blue, linktoc=page, bookmarks=false, pdfstartview={FitH}, pdfborder={0 0 0.0 [3 3]}]{hyperref} 
\usepackage[usenames]{color}
\hypersetup{pdfborder=0 0 0,colorlinks=true,citecolor=blue,linkcolor=blue}
\usepackage[final]{pdfpages}
\usepackage{epstopdf}
\usepackage{subfigure}
\usepackage{units}
\usepackage{url}

\def\sint{\ifmmode{- \!\!\!\!\!\! \int}
	\else{\hbox{$- \!\!\!\! \int \ $}}\fi}

\begin{document}

	\title{Anisotropic magneto-resistance in MgO-based magnetic tunnel junctions induced by spin-orbit coupling}

	\author{Hui-Min Tang}
	\affiliation{School of Physical Science and Technology, Guangxi Normal University, Guangxi 541001, China}
	\author{Shi-zhuo Wang}
	\affiliation{School of Physics and Electronic Engineering, Zhengzhou University of Light Industry,
		Zhengzhou 450002, China}
	\author{Xin-tao Jia}
	\affiliation{School of Physics and Electronic Information Engineering, Henan Polytechnic University, Jiaozuo 454000, China}
	
	\begin{abstract}
	We performed a first-principles study of the tunneling anisotropic magneto-resistance (TAMR) in Ag(Ir,Pt)$/$MgO$/$Fe junctions. Enhanced TAMR with ideal and skewed fourfold angular dependence is found in-plane and out-of-plane TAMR of the system, respectively, which shows simple barrier thickness dependency with number around 10\% in some junctions. The complex angular dependency of the interfacial resonant states due to the spin-orbit coupling should be responsible to the complex and enhanced TAMR found in these junctions.
	\end{abstract}

\maketitle

\section{Introduction} 
Anisotropic magneto-resistance (AMR) effect, which is around several percent generally, found in the ferromagnetic alloys\cite{mcguire1975anisotropic} and ferromagnet$/$normal metal (FM$/$NM) interface was well understand for a long time. However, the AMR effect in the ferromagnetic alloys and a single FM$/$NM interface is too weak to function well for spintronic applications. To enhance the AMR effect in the FM$/$NM interface, functional layer such as insulator barrier has been introduced.\cite{PhysRevLett.93.117203,PhysRevB.93.024419,Pham2016,PhysRevLett.110.037202,PhysRevB.88.054407,huangfu2012,PhysRevLett.95.086604,matos2009anisotropic,PhysRevLett.98.046601,herve2018large,PhysRevLett.99.226602} AMR effect is prevailing in the antiferromaget, also. AMR larger than 100$\%$ at 4 $\unit{K}$ was reported in an NiFe$/$IrMn$/$MgO$/$Pt spin valve structure, which is reduced by a few percent at higher temperature unfortunately.\cite{Park2011}

The weak spin-orbit coupling (SOC) in the ferromagnetic alloys and magnet$/$normal metal interface leads to weak magnetoresistance therein. Enhanced structural asymmetry by introducing the inserting insulator layer in the FM$/$NM interface can increase the AMR effect considerably.\cite{PhysRevLett.93.117203,PhysRevB.93.024419,Pham2016,PhysRevLett.110.037202,PhysRevB.88.054407,huangfu2012,PhysRevLett.95.086604,matos2009anisotropic,PhysRevLett.98.046601,herve2018large,PhysRevLett.99.226602} to further increase the AMR effect, well-designed structure with resonant surface states are introduced,\cite{PhysRevLett.98.046601,herve2018large} which is the key to understand the huge tunnel magentoresistance (TMR) in the Fe$/$Vacuum$/$Fe junction\cite{PhysRevB.73.180402}. Furthermore, the localized interfacial states\cite{PhysRevLett.99.226602} and resonant level\cite{PhysRevB.77.165412} can affect the AMR effect also. A complex bias-dependent AMR is recent found in FeCo$/$MgO$/$FeCo and FeCo$/$Al$_2$O$_3$$/$FeCo ferromagnetic tunnel junctions (MTJs),\cite{PhysRevLett.99.226602} which is related with the coupling of the resonant states with the interfacial density of states.

In this letter, we carry first-principles calculations on spin transport on the NM$/$MgO$/$Fe MTJs with to investigate the tunneling AMR (TAMR), where NM stands for Ag, Pt, and Ir. A ideal fourfold angular dependence is found in the in-plane TAMR of the system, while a screwed fourfold angular dependence in out-of-plane TAMR. The complex angular dependency of the interfacial resonant states due to the spin-orbit coupling should be responsible to the complex and enhanced TAMR found in these junctions. This article is organized as follows. In Sec. \uppercase\expandafter{\romannumeral2}, we introduce two configurations of TAMR: the in-plane and out-of-plane configurations. In Sec. \uppercase\expandafter{\romannumeral3}, we present our results of TAMR in NM$/$MgO$/$Fe MTJs. Section \uppercase\expandafter{\romannumeral4} is our summary.

\section{Two different configurations of TAMR} 
\begin{figure}
 	\includegraphics[width=12cm]{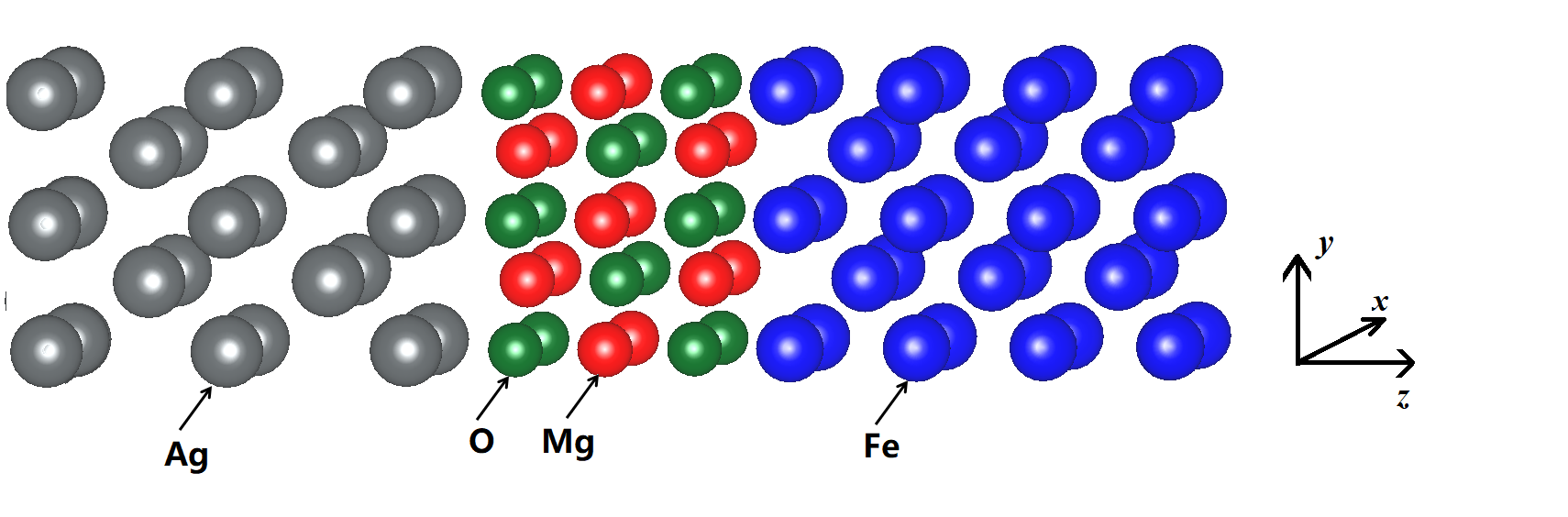}
 	\caption{Schematic Ag$/$MgO-cap(001)$/$Fe MTJs with three MgO monolayers. The red and green atoms in the scattering region (MgO) denote Mg and O, the gray atoms on the left leads denote Ag, and the blue atoms denote Fe. The transport along to the z direction. }
 	\label{structure}
 \end{figure}

According to the epitaxial structure of the junctions, two configurations should be considered for investigating the TAMR effect, that is the in-plane and out-of-plane configurations. Considering a NM$/$MgO$/$Fe junction, as shown in Fig. \ref{structure}, the in-plane and out-of-plane configurations refers to the Fe magnetization rotated in the $x-y$ and $x-z$ planes, respectively. The in-plane and out-of-plane TAMR are defied as\cite{PhysRevB.79.155303,PhysRevB.80.045312}
\begin{eqnarray}
\rm{TAMR}^{in}_{[x]}(\theta=90^\circ,\phi)=\frac{R(\theta,\phi)-R(\theta,0)}{R(\theta,0)},
\label{in-plane TAMR} \\
\rm{TAMR}^{out}_{[x]}(\theta,\phi=0)=\frac{R(\theta,\phi)-R(0,\phi)}{R(0,\phi)},
\label{eq. TAMR}
\end{eqnarray}
where $R(\theta,\phi)$ denotes the tunneling magneto-resistance for the magnetization oriented along the direction defined by the unit vector $\mathbf{m}=(sin\theta cos\phi,sin\theta sin\phi,cos\theta)$.

The form of angular dependence is an important property of the TAMR. Previously theoretical model suggest that the angle dependent is determined by the specific form and symmetry properties of the interface-induced SOC field.\cite{PhysRevB.80.045312}

\section{TAMR in NM$/$MgO$/$Fe MTJs}  
We consider a NM$/$MgO$/$Fe MTJs, as shown in Fig. \ref{structure}. In our calculation, we choose bcc Fe as one lead with crystal constant a$_{Fe}$=2.866 $\rm\AA$, the crystal MgO is reduced 4\% and rotated 45 degree to match the bcc Fe. The NM we consider fcc Ag, Ir and Pt, the crystal constant is a$_{Ag}$=4.053 $\rm\AA$, a$_{Ir}$=3.839 $\rm\AA$ and a$_{Pt}$=3.9242 $\rm\AA$. The in-plane lattice constant of the NM is matched with MgO, and the interlayer spacing between the NM monolayer and the MgO surface layer was 2.520 $\rm\AA$, 2.305 $\rm\AA$ and 2.364 $\rm\AA$ for Ag, Ir and Pt above the O site to ensure the space full filling. We used a 400$\times$400 k-point mesh in the full two-dimensional Brillouin zone (BZ) to ensure numerical convergence. The scattering matrix was obtained using a first-principles wave function method with tight-binding linearized muffin-tin orbitals (TB-LMTO)\cite{PhysRevB.12.3060,PhysRevB.34.5253} include SOC\cite{PhysRevLett.105.236601}.

 \begin{figure}
 	\subfigure{\includegraphics[width=13cm]{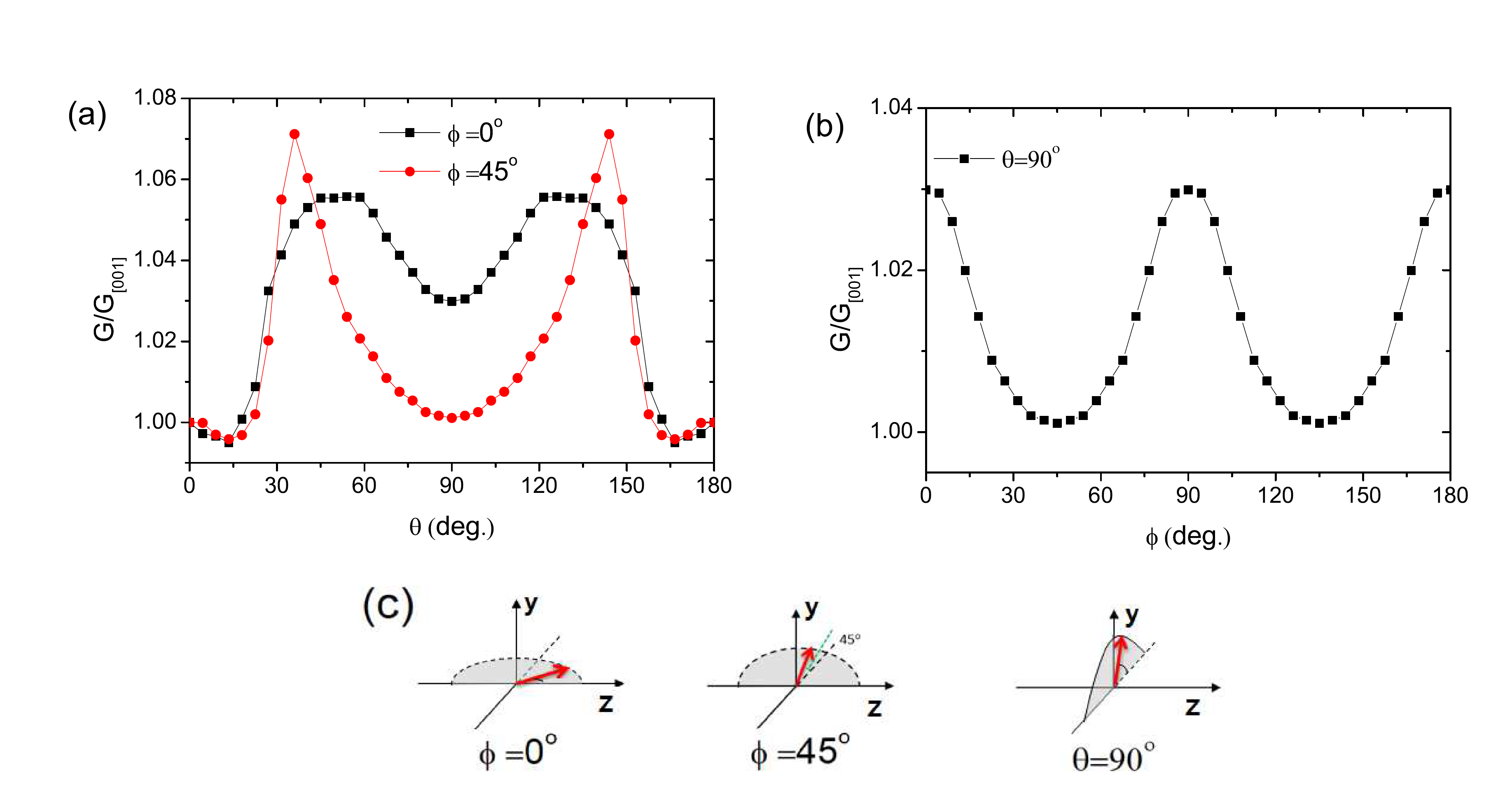}} 
 	\caption{Angular dependence of tunneling conductance of Ag$/$MgO(3\,ML)$/$Fe MTJs with clean interface. The data are normalized by conductance G$_{[001]}$ where the current was along [z] direction. (a) The magnetization is rotation in $\phi=0$ (xz) and $\phi=45^\circ$ plane. (b) The magnetization is rotation in $\theta=90^\circ$ (xy) plane. (c) A schematic picture of the rotating magnetization of FM.}
 	\label{Ag-AMR}
 \end{figure}

Fig. \ref{Ag-AMR} shows the angular dependence of tunneling conductance of Ag$/$MgO(3\,ML)$/$Fe MTJs with clean interface. The conductances are normalized by conductance G$_{[001]}$ where the current was along [001]. The magnetization is rotation in $\phi=0$ (xz), $\phi=45^\circ$, and $\theta=90^\circ$ (xy) plane, the corresponding schematic diagram was shown in Fig. \ref{structure} (c).

A fourfold angular dependence was shown in Fig. \ref{Ag-AMR} (a), with peaks at $\phi=50^\circ, 130^\circ$, valleys at $\phi=17.5^\circ, 162.5^\circ$, when the magnetization is rotation in xz plane, and with peaks at $\theta=35^\circ, 145^\circ$, valleys at $\theta=17.5^\circ, 162.5^\circ$ of $\phi=45^\circ$ plane. The magnetization is rotation in $\theta$=90$^{\circ}$ (xy) plane, we found the relation $G/G_{[001]}$ is larger than 1, this results shows that conductance which the magnetization perpendicular to current is large than that in the parallel direction, this is consistent with the results of AMR.\cite{McGuire1975}

 \begin{figure}
 	\includegraphics[width=13cm]{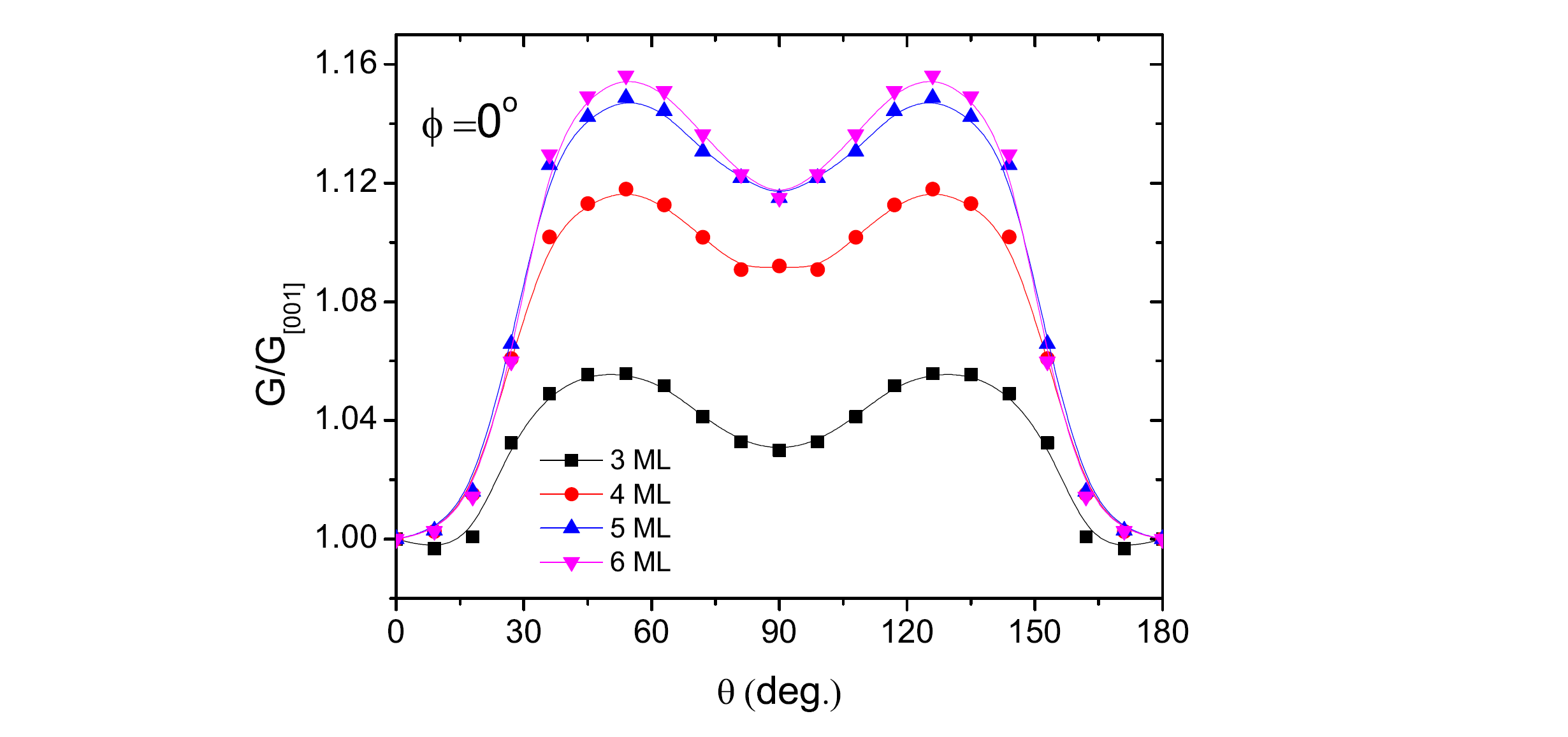}
 	\caption{Angular dependence of tunneling conductance of Ag$/$MgO(x\,ML)$/$Fe MTJs with clean interface. x is range from 3\,ML to 6\,ML. The data are normalized by conductance G$_{[001]}$ where the current was along [z] direction. The magnetization is rotation in xz plane.}
 	\label{L-MgO}
 \end{figure}

To understand how the thickness of MgO barrier affected, we studied angular dependence of tunneling conductance as a function of MgO thickness of Ag$/$Mg$/$Fe MTJs with clean interface, as shown in Fig. \ref{L-MgO}. The thickness of the MgO layer ranging from 3 ML to 6 ML. Although the tunneling conductance decreases, the relation $G/G_{[100]}$ increases, and therefore the AMR increases.

\begin{table}[h]
	\caption{TAMR of Ag$/$MgO(3\,ML)$/$Fe MTJs with clean interface, n is the number of layer of MgO. The second low is the rotation plane of the magnetization $\mathbf{M}$, the unit of conductance is $ (10^{-3}e^2/h)$. $\rm{TAMR}^{out}_{[x]}=\frac{G_{max}-G_{[001]}}{G_{[001]}}$ as shown in eq. \ref{eq. TAMR}.}
	\begin{center}
		\begin{tabular*}{8cm}[c]{@{\extracolsep{\fill}}cc|c|cccc}\hline\hline	
			& n & $\mathbf{M}$  &$G_{[001]}$  & $G_{max}$ & $\rm{TAMR}^{out}_{[x]}(\%)$ \\
			\hline
			& 3     & $x\text{-}y$     & 12.84  & 13.22 & 2.959   \\		
			&       & $y\text{-}z$     & 12.84  & 13.56 & 5.607   \\
			&       & $\phi=45^\circ$  & 12.84  & 13.75 & 7.087   \\
			\hline
			& 4     &$y\text{-}z$      & 2.015  & 2.253 & 11.81   \\
			& 5     &$y\text{-}z$      & 0.402  & 0.462 & 14.92   \\
			& 6     &$y\text{-}z$      & 0.085  & 0.098 & 15.29   \\
			\hline\hline
		\end{tabular*}
	\end{center} \label{tab:Ag-TAMR}
\end{table}

Table \ref{tab:Ag-TAMR} shows the TAMR of Ag$/$MgO(3\,ML)$/$Fe MTJs with clean interface, n is the number of layer of MgO. We list the results of the maximum of tunneling conductance, and the value of $\rm{TAMR}^{out}_{[x]}=\frac{G_{max}-G_{[001]}}{G_{[001]}}$, as shown in eq. \ref{eq. TAMR}. The max value of TAMR is 7.087\% when the magnetization is rotation in $phi=45^{\circ}$ plane. As the thickness of MgO barrier increase, so does the TAMR. The TAMR is large to 15.29\% with 6 ML MgO barrier (about 1.2 nm). The AMR of sharvin Fe is very small, about 0.18\%.

  \begin{figure}
  	\includegraphics[width=13cm]{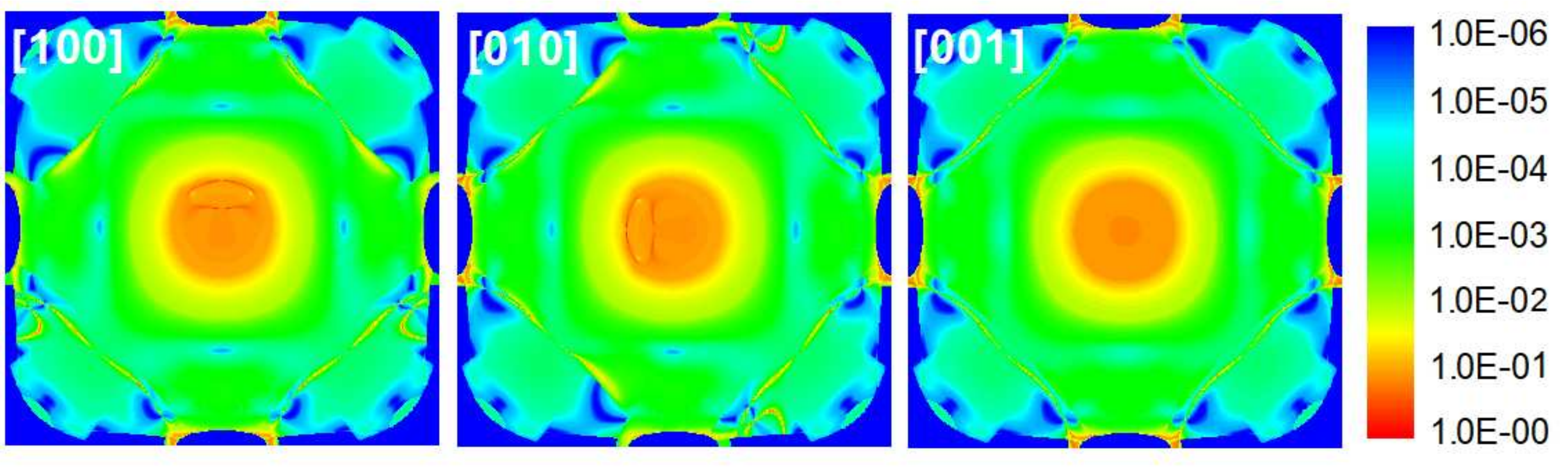}
  	\caption{k$_{\parallel}$ resolved electron transmission probability for Ag$/$MgO(3\,ML)$/$Fe MTJs with clean interface. The transport is along [001] direction, and the magnetization of Fe is along [100], [010] and [001] axis.}
  	\label{kmesh}
  \end{figure}

To understand the angular dependence of conductance, we plot the k$_\parallel$ resolved transmission at the Fermi energy for the epitaxial Ag$/$MgO(3\,ML)$/$Fe MTJs with clean interface, as shown in Fig. \ref{kmesh}. The logarithm function has been applied to the transmission coefficient, and the red (blue) color represents the high (low) transmission probability. We observed that the transmission channel is mainly come from the k$_\parallel$ points near to the $\Gamma$ point, and there are some high transmission channel near the edge of BZ. The BZ is fourfold symmetry of [001] direction, and twofold symmetry of [100] ([010]) relative to k$_y$=0 (k$_x$=0). There are a few hot spots near the $\Gamma$ point of [100] and [010], that is the origin of the G$_{[001]}$ is smaller than G$_{[100]}$ and G$_{[010]}$, which the tunneling conductance when the magnetization perpendicular to current is large than that in the parallel direction.

  \begin{figure}
  	\includegraphics[width=13cm]{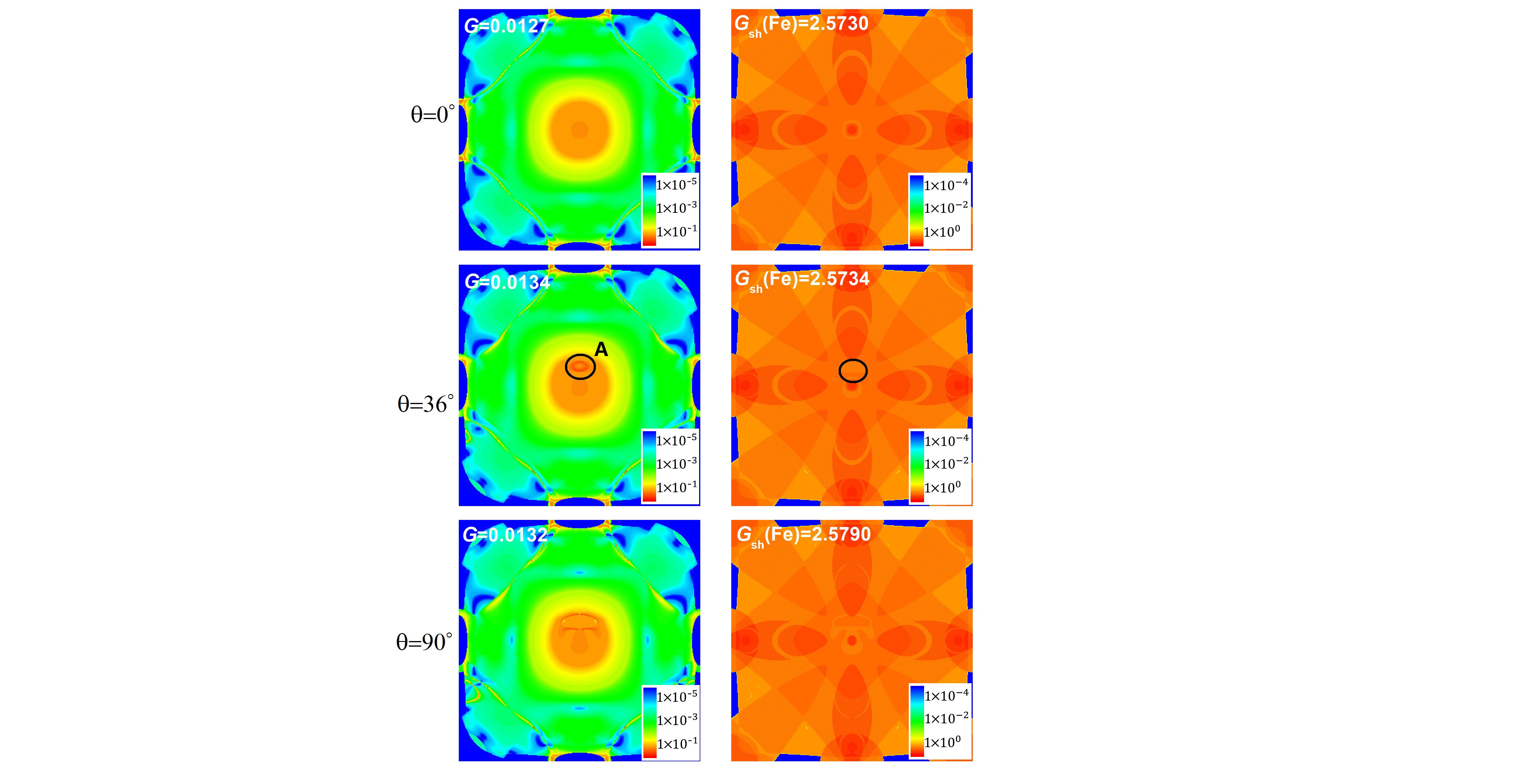}
  	\caption{k$_{\parallel}$ resolved electron transmission probability and sharvin conductance for Ag$/$MgO(3\,ML)$/$Fe MTJs with clean interface. The transport is along [001] direction, the magnetization is rotation in $\phi=0$ (xz) plane.}
  	\label{sharvin}
  \end{figure}

\begin{figure}
   	\includegraphics[width=13cm]{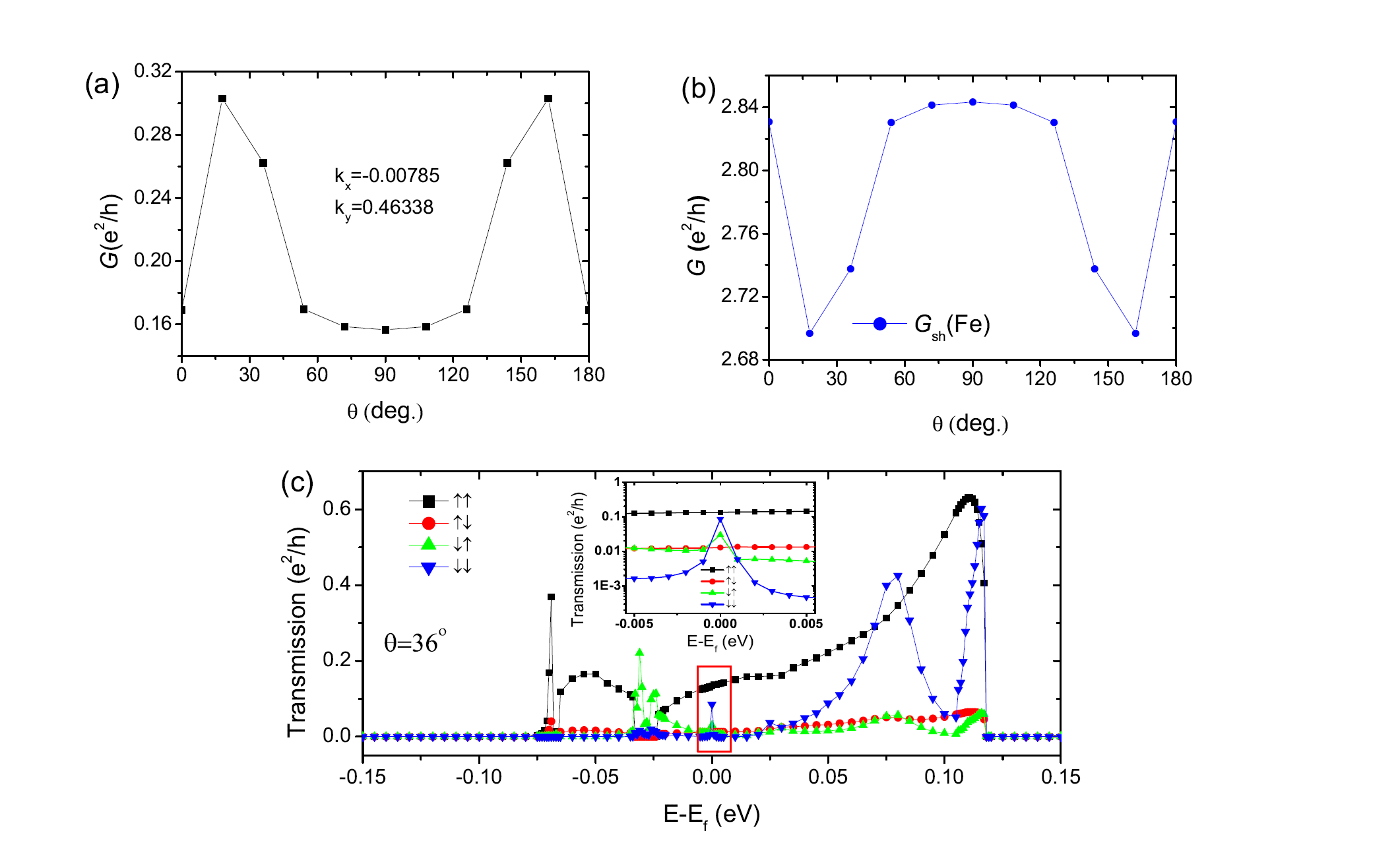}
   	\caption{(a) Angel dependent conductance and (b) sharvin conductance of Fe as a function of $\theta$ in Ag$/$MgO(3\,ML)$/$Fe MTJs with clean interface at high transmission points near to the $\Gamma$ point, which k$_x$a$_0$=-0.00785 and k$_y$a$_0$=0.46338, where a$_0$=0.286 nm is the lattice constant of Fe. (c) Energy-dependent transmission of Ag$/$MgO(3\,ML)$/$Fe MTJs at the same k points.}
   	\label{Fe-sharvin-angle}
   \end{figure}

We plot the k$_\parallel$ resolved transmission and sharvin conductance of Fe electrode at the Fermi energy for the epitaxial Ag$/$MgO(3\,ML)$/$Fe MTJs with clean interface, with the magnetization is rotation in $\phi=0$ (xz) plane. We found large transmission channel caused by the resonant states at those k$_\parallel$ points near the $\Gamma$ point, when $\theta=36^\circ, 90^\circ$, as shown in black cycle (points A). The sharvin conductance of Fe changes at  $\theta=36^\circ$ respect to $\theta=0$, which is effected by the SOC, as shown in Fig. \ref*{Fe-sharvin-angle}. The coupling of SOC and resonant state is the source of the complex angular dependence TAMR effects.

\begin{figure}
	\centering
	\includegraphics[width=13cm]{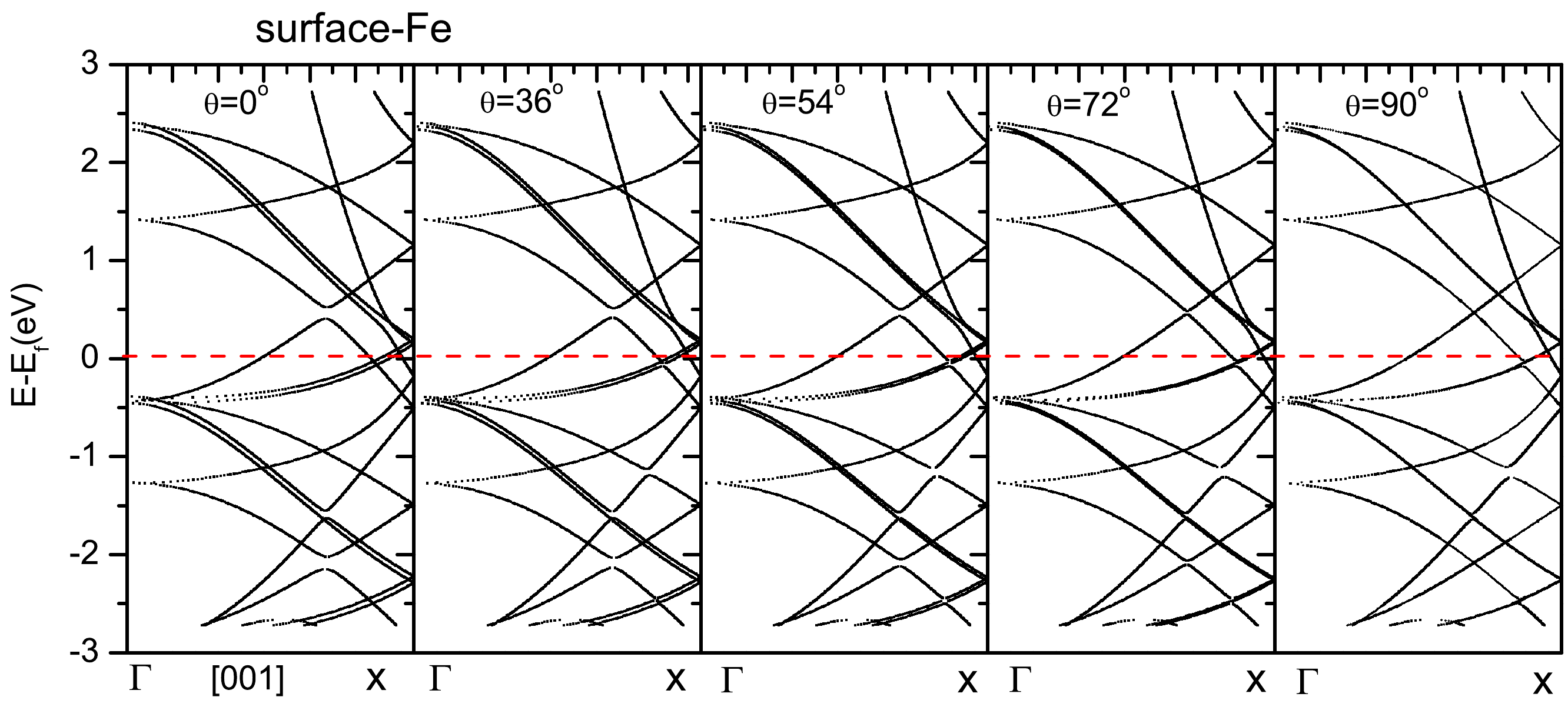}
	\caption{Band structure of the interface Fe atom at different angles along the transport direction in Ag$/$MgO(3ML)$/$Fe MTJs whth clean interface. $\phi=0$ is the magnetization is rotate in $xz$ plane. $\phi=0$ ($\phi=0$) denotes the magnetization is parallel (vertical) to current. }
	\label{fig:band-fe}
\end{figure}

Band structure of the interface Fe atom along the transport direction with different angles are analyzed in the following to further understand the angle dependent of tunneling conductance, as shown in Fig. \ref*{fig:band-fe}. Spin-orbit coupling opens the degenerate band at the parallel direction. When the magnetization deviates from the direction of the current, the spin-orbit coupling effect becomes smaller, and the previously opened degenerate states overlap again, resulting in the difference of conductance.

\begin{figure}
	\centering
	\includegraphics[width=13cm]{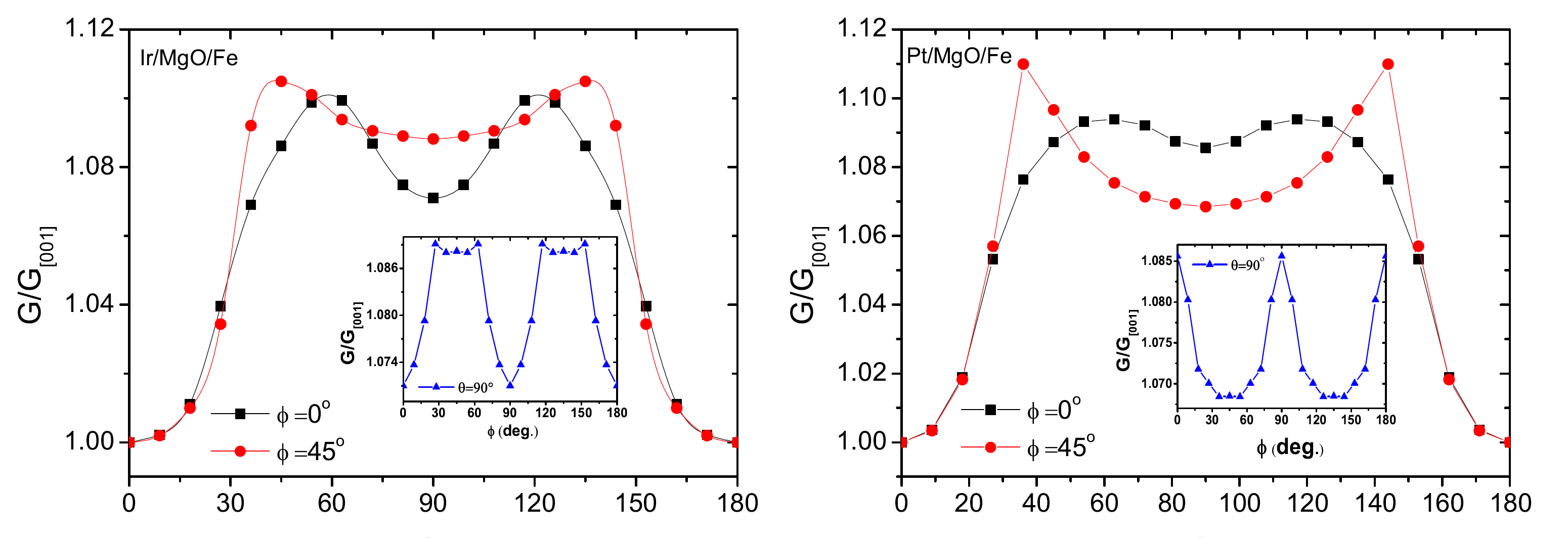}
	\caption{Angular dependence of tunneling conductance of NM$/$MgO(3\,ML)$/$Fe MTJs with clean interface. The NM we choose Ir and Pt.  The data are normalized by conductance G$_{[001]}$ where the current was along [z] direction. The magnetization is rotation in $\phi=0$ (xz) and $\phi=45^\circ$ plane. The inset picture is the angular dependence of tunneling conductance when the magnetization is rotation in $\theta=90^\circ$ (xy) plane. }
	\label{fig:IrPt}
\end{figure}

Perpendicular magnetic anisotropy (PMA) has been demonstrated that is critically dependent on the capping layer materials in
CoFeB$/$MgO-based structures.\cite{Jeon2015,Liu2012,Young2014,srep18173} Thus, we estimate the effect of NM capping layer of NM$/$MgO$/$Fe MTJs. Fig. \ref{fig:IrPt} shows the angular dependence of tunneling conductance of Ir$/$MgO$/$Fe and Ir$/$MgO$/$Fe MTJs with clean interface. A fourfold angular dependence was shown in the results both Ir and Pt, the value of AMR increases with increasing thickness of the MgO barrier, and large to 15.8\% (18.5\%) at 6\,ML MgO barrier of Pt$/$MgO$/$Fe (Ir$/$MgO$/$Fe).

\section{Summary}  
We performed a first-principles study of the tunneling anisotropic magneto-resistance (TAMR) in Ag$/$MgO$/$Fe magnetic tunnel junctions. We find a fourfold symmetry of angular dependence of conductance in these systems. The peaks at $\phi=50^\circ, 130^\circ$, valleys at $\phi=17.5^\circ, 162.5^\circ$, when the magnetization is rotation in xz plane, and with peaks at $\theta=35^\circ, 145^\circ$, valleys at $\theta=17.5^\circ, 162.5^\circ$ of $\phi=45^\circ$ plane. The TAMR effect derives from the anisotropy in the interface density of states of the majority band due to mixing with a resonant state via spin-orbit coupling. The value of TAMR is effect by the thickness of MgO barrier.

\section{acknowledgments}
We gratefully acknowledge financial support from the National Natural Science Foundation of China (Grants No. 11804062), and the financial from the Natural Science Foundation of Guangxi through Grant No. 2018GXNSFAA138160.

\bibliographystyle{apsrev}
\bibliography{AMR}
\end{document}